\documentclass[%
reprint,
 amsmath,amssymb,
 aps,
]{revtex4-1}

\usepackage{graphicx}
\usepackage{dcolumn}
\usepackage{bm}
\usepackage{comment}
\usepackage[flushleft]{threeparttable}

\begin{document}


\title{Multi-objective and categorical global optimization of photonic structures based on ResNet generative neural networks}

\author{Jiaqi Jiang}
\author{Jonathan A. Fan}
\affiliation{Stanford University, Department of Electrical Engineering, Stanford, CA, United States}%

\date{\today}

\begin{abstract}
We show that deep generative neural networks, based on global topology optimization networks (GLOnets), can be configured to perform the multi-objective and categorical global optimization of photonic devices.  A residual network scheme enables GLOnets to evolve from a deep architecture, which is required to properly search the full design space early in the optimization process, to a shallow network that generates a narrow distribution of globally optimal devices.  As a proof-of-concept demonstration, we adapt our method to design thin film stacks consisting of multiple material types.  Benchmarks with known globally-optimized anti-reflection structures indicate that GLOnets can find the global optimum with orders of magnitude faster speeds compared to conventional algorithms.  We also demonstrate the utility of our method in complex design tasks with its application to incandescent light filters.  These results indicate that advanced concepts in deep learning can push the capabilities of inverse design algorithms for photonics.

\end{abstract}

\maketitle


\section{Introduction}
Inverse algorithms are amongst the most effective methods for designing efficient, multi-functional photonic devices \cite{molesky2018inverse,ReviewWerner, fan_2020}.  It remains an open question how to select and implement a design algorithm, and over the last few years, much research has been focused on deep neural networks as inverse design tools \cite{jiang2020deep, yao2019intelligent, so2020deep}.  Many of these demonstrations are based on the generation of a training set, consisting of device geometries and their optical responses, and modeling these data using discriminative  \cite{peurifoy2018nanophotonic, liu2018training} or generative \cite{ma2019probabilistic, jiang2019free, liu2018generative, wen2020robust} neural networks.  These methods have proven to be capable of producing high speed surrogate solvers and can perform inference-type tasks with training data.  When the training data are curated using advanced gradient-based optimization methods, such as the adjoint variables \cite{hughes2018adjoint, sell2017large, piggott2015inverse, Phan2019, JianjiAnnalen} or objective-first methods \cite{lu2012objective}, the networks can learn to generate high performing, freeform photonic structures.


To perform global optimization, alternative approaches are required that do not depend on interpolation from a training set.  The reason is because the design space is non-convex and contains multiple local optima, and even devices based on advanced gradient-based optimization methods cannot help a neural network search for the global optimum.  In this vein, global optimization networks (GLOnets)  have been developed to perform the non-convex global optimization of freefrom photonic devices \cite{jiang2019global, jiang2019simulator}.  GLOnets are gradient-based optimizers that do not use a training set but instead combine a generative neural network with an electromagnetic simulator to perform population-based optimization. The evolution of the generated device distribution is driven by both figure-of-merit values (i.e., efficiencies) and gradients for devices sampled from the generative network. Initial implementations of GLOnets were configured for single-objective problems with binary design variables, such as the maximization of deflection efficiency for a normally incident beam in a metagrating comprising silicon nanostructures.  ``Single-objective" refers to the optimization of a system operating with one conditional parameter, in this case a system with fixed incidence beam angle, and ``binary" refers to silicon and air as our design materials.  

A more general formulation of the problem that captures the design space of many photonic technologies is multi-objective, categorical optimization with more than two design materials.  ``Multi-objective" refers to the optimization of a system operating involving more than one objective function to be optimized simultaneously, such as a metagrating operating over a range of incident beam angles, and ``categorical" refers to design variables that have two or more categories without intrinsic ordering, such as multiple material types.  In this study, we show that GLOnets can be configured as a multi-objective, categorical global optimizer, and we we adapt GLOnets to optimize thin films stacks to demonstrate the capabilities of our algorithms.  Thin film stacks are an ideal model system for multiple reasons.  First, the design problem is multi-objective, as devices are typically configured for a range of incident wavelengths, angles, and polarizations.  Second, the design problem is categorical, as individual layer materials are chosen from a library of materials.  Third, thin film stacks are a well established technology, and there are a number of pre-existing studies that enable proper benchmarking of algorithm performance \cite{shi2017optimization, azunre2019guaranteed, wang2020automated}.

Thin film stacks have been widely used in many optical systems including passive radiative coolers \cite{raman2014passive}, efficient solar cells \cite{li2017comprehensive, lenert2014nanophotonic}, broadband spectral filtering \cite{shen2014optical,cao2016toward}, thermal emitters \cite{ilic2016tailoring}, and spatial multiplexing filters \cite{gerken2003wavelength}.  The materials and thicknesses of thin film layers have to be carefully optimized to achieve the desired transmission and reflection proprieties across a broad wavelength and angular bandwidth.  Design methods based on physical intuition result in limited performance, and they are generally difficult to scale to aperiodic thin film stacks comprising many layers.  To address these limitations, various global optimization approaches have been explored, including the Monte Carto approach \cite{wild1986thin}, particle swarm optimization \cite{rabady2014global}, needle optimization \cite{tikhonravov1996application, pervak20071, tikhonravov2007optical}, and the memetic algorithm \cite{shi2017optimization}. These methods are all derivative-free global optimization algorithms that search the design space through the evaluation of a batch of samples without any gradient calculations, limiting their ability to reliably solve for the global optimum.


\section{Method}

\begin{figure}[ht!]
    \centering
    \includegraphics[width = 0.9\linewidth]{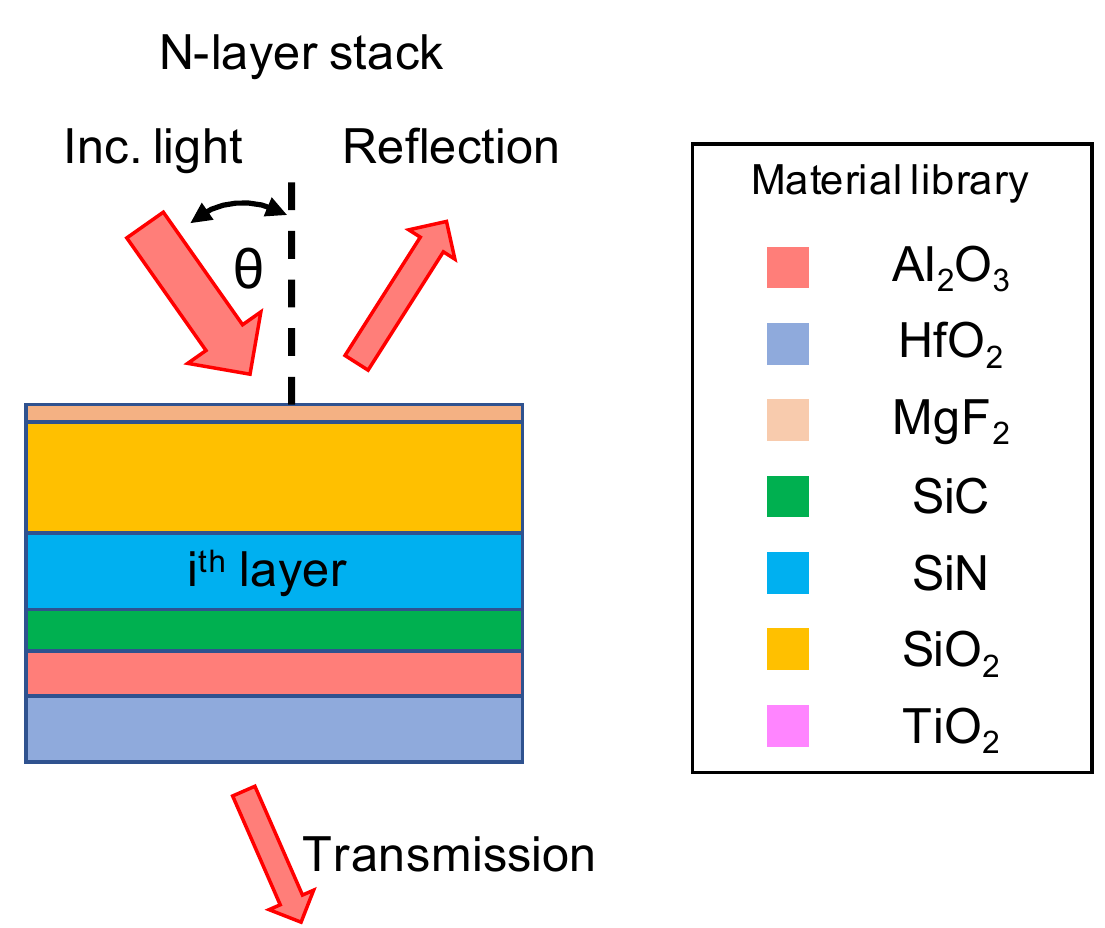}
    \caption{Schematic of the N-layer thin film stack system.  The refractive index and thickness of each layer are optimized to produce a desired reflection profile, and the composition of each layer is constrained to index values specified in a material library.
    }
    \label{fig:1}
\end{figure}

We consider the design of $N$-layer thin film stacks each comprising an isotropic material specified from a material library (Figure \ref{fig:1}).  The refractive indices of the total stack are denoted as a vector $\mathbf{n}(\lambda) = (n_1 (\lambda), n_2(\lambda), \cdots, n_N(\lambda))^T$, where each index term is a function of wavelength to account for dispersion and the values can be real or complex-valued without loss of generality.  The thin film stack thicknesses are $\mathbf{t} = (t_1, t_2, \cdots, t_N )^T$.  The material library consists of $M$ material types and their refractive indices are represented as $\{m_1 (\lambda), m_2(\lambda), \cdots, m_M(\lambda)\}$.  

The optimization problem is posed as finding the proper $\mathbf{n}$ and $\mathbf{t}$ that produces the desired reflection characteristics over a given wavelength bandwidth, incident angle range, and incident polarization:

\begin{equation}
    \{\mathbf{n}^*, \mathbf{t}^*\} = \arg\min_{\{\mathbf{n}, \mathbf{t}\}} \sum_{\lambda, \theta, \text{pol}} \left( \mathcal{R}(\mathbf{n}, \mathbf{t}|\ \lambda, \theta, \text{pol}) -  \mathcal{R}^*( \lambda, \theta, \text{pol})\right)^2
\end{equation}

The desired reflection spectrum is denoted as $\mathcal{R}^*( \lambda, \theta, \text{pol})$, and $\{\mathbf{n}^*, \mathbf{t}^*\}$ are the corresponding global optimal refractive indices and thicknesses.  This optimization problem can be readily cast as the minimization of the objective function: $O(\mathbf{n}, \mathbf{t})=\sum_{\lambda, \theta, \text{pol}} \left( \mathcal{R}(\mathbf{n}, \mathbf{t}|\ \lambda, \theta, \text{pol}) -  \mathcal{R}^*( \lambda, \theta, \text{pol})\right)^2$ .  $\mathbf{n}$ are categorical variables because the index values are chosen from a material database,  while $\mathbf{t}$ can span a continuous set of values and is a continuous variable.

\subsection{Transfer matrix method solver}

A principle requirement of any gradient-based optimizer is a method to calculate local gradients. For thin film stacks, these gradients indicate how perturbations to the refractive indices and thicknesses of the device can best reduce the objective function.  In prior implementations of GLOnets, local gradients were calculated using the adjoint variables method, in which a forward and adjoint simulation are calculated using a conventional electromagnetic solver \cite{jiang2019global, jiang2019simulator}.  

While the adjoint variables method provides a general formalism to calculating local gradients using any conventional solver, we pursue an alternative approach based on the transfer matrix method (TMM), which is a fully analytic and high speed solver for thin film systems.  In particular, we program a TMM solver within the automatic differentiation framework in PyTorch \cite{paszke2019pytorch}, which allows gradients to be directly calculated using the chain rule.  Automatic differentiation is the basis for calculating gradients during backpropagation in neural network training, and it generally applies to any algorithm that can be described by a differentiable computational graph.  Recently, it was implemented in finite-difference time domain (FDTD) and finite-difference frequency domain (FDFD) simulators \cite{hughes2019forward, minkov2020inverse}.  Compared to generalized differentiable electromagnetic solvers, such as these FDTD and FDFD implementations, our analytic TMM-based algorithms are faster without loss of accuracy because the thin films are described as layers instead of voxels.

\subsection{Res-GLOnet algorithm}

\begin{figure*}[ht!]
    \centering
    \includegraphics[width = 0.9\linewidth]{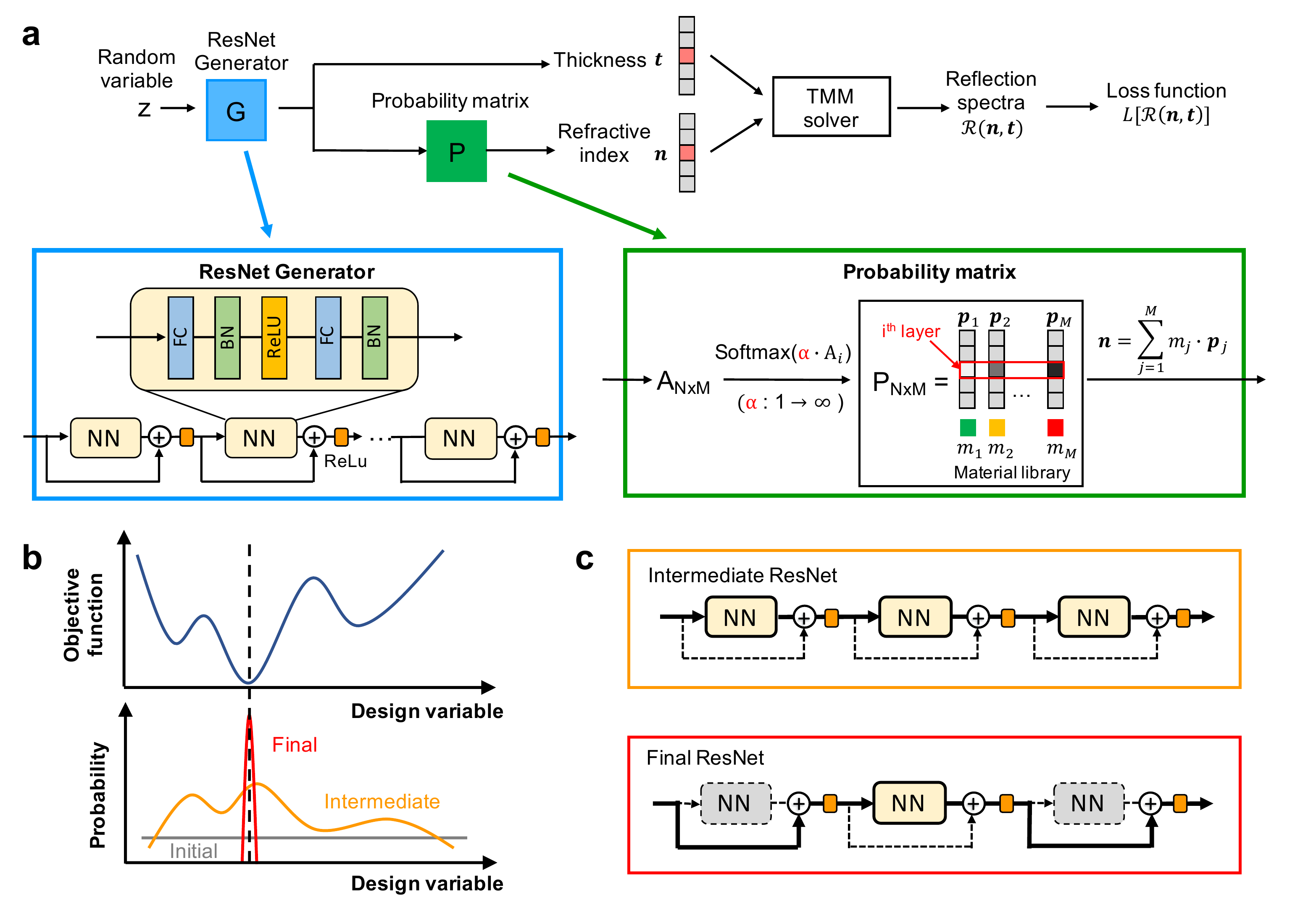}
    \caption{Thin film global optimization with Res-GLOnet. (a) Schematic of the Res-GLOnet.  A ResNet generator maps a uniformly distributed random variable to a distribution of devices, which are then evaluated with a transfer matrix method solver and used to evaluate the loss function.  A probability matrix pushes the continuous generated device indices $\mathbf{n}$ to discrete values.  (b) Evolution of the generated device distribution over the course of network training.  The network initially samples the full design space and converges to a narrow distribution centered around the global minimum of the objective function.  (c) During training, the network operates as a deep architecture with little impact from the skip connections (Intermediate ResNet).  Near training completion, the network evolves to a shallow architecture with large impact from the skip connections (Final ResNet).  Bold and dashed lines indicate large and small contributions to the network architecture, respectively.}
    \label{fig:2}
\end{figure*}

A schematic of GLOnets configured for our thin film stack system is outlined in Figure \ref{fig:2}a.  We term this GLOnets variant as Res-GLOnets because the generator has a residual network architecture that includes skip connections between layers (blue box inset), which will be discussed in a later section. First, a generative neural network $G$ with trainable weights $\phi$ produces a distribution of thin-film stack configurations.  The input to the generator is a uniformly distributed random vector $\mathbf{z} \sim U(0,1)$, so that the generator can be regarded as a function that maps the uniform distribution to a complex distribution of thin-film stack configurations, $G_{\phi}: U(0,1) \rightarrow P_{\phi}(\mathbf{n}, \mathbf{t})$.  Different samplings of the input random variable $\mathbf{z}^{(k)}$ map onto different device refractive index and thickness configurations within $P_{\phi}(\mathbf{n}, \mathbf{t})$, denoted as $\{\mathbf{n}^{(k)},\mathbf{t}^{(k)}\} = G_{\phi}(\mathbf{z}^{(k)})$.  The generated $\mathbf{n}$ from the network do not take categorical values from the materials library but are relaxed to be continuous variables, to stabilize the optimization process.  These $\mathbf{n}$ are further processed by a probability matrix to enforce the categorical value constraint, which is discussed in the next section.  After processing, the reflection spectra of the generated devices, $ \mathcal{R}(\mathbf{n}^{(k)}, \mathbf{t}^{(k)}|\ \lambda, \theta, \text{pol})$,  are calculated using the TMM solver.

The optimization objective, or the loss function, for GLOnet is defined as:
\begin{align}
     L &= \mathbb{E}\left[ \exp{\left(-\frac{O(\mathbf{n}, \mathbf{t})}{\sigma} \right)} \right]  \label{eq2} \\ 
     &=  \int \exp{\left(-\frac{O(\mathbf{n}, \mathbf{t})}{\sigma} \right)} P_{\phi}(\mathbf{n}, \mathbf{t})\ d\mathbf{n} d\mathbf{t} \label{eq3}\\ 
     &= \int  \exp{\left(-\frac{O(G_{\phi}(\mathbf{z}))}{\sigma} \right)} P(\mathbf{z})\ d\mathbf{z} \label{eq4} \\
     &\approx \sum_{k=1}^K \exp{\left(-\frac{O(\mathbf{n}^{(k)}, \mathbf{t}^{(k)})}{\sigma} \right)}
\end{align}
$\sigma$ is a hyperparameter.  These equations follow the derivation of the GLOnet formalism described in Ref. \cite{jiang2019simulator}.  To train the generative network and update its weights in a manner that improves the mapping of $\mathbf{z}$ to devices, the gradient of the loss function with respect to the neuron weights, $\nabla_{\phi} L$, is calculated by backpropagation. 

A schematic of the evolution of the generative network over the course of network training is outlined in Figure \ref{fig:2}b.  Initially, the generator has no knowledge about the design space and outputs a broad distribution of devices spanning the full design space.  Over the course of network training, the distribution of generated devices narrows and gets biased towards design space regions that feature relatively small objective function values.  Upon the completion of network training, the distribution of generated thin film stack configurations converges to a narrow distribution centered around the global optimum.


\subsection{Enforcing categorical constraints}

To update the weights in the generative network during backpropagation, the chain rule is applied to the entire computation graph of the Res-GLOnet algorithm.  One required step is the calculation of the gradient of the reflection spectrum with respect to the refractive indices, $\frac{d\mathcal{R}}{d\mathbf{n}}$. If the refractive indices of thin-film stacks outputted by the generator are directly treated as categorical variables, $\mathbf{n}$ is not a continuous function and the gradient term above cannot be calculated.  

To overcome this difficulty, we propose a reparameterization scheme in which the generated $\mathbf{n}$ are relaxed to take continuous values and are then processed in a manner that supports convergence to categorical variable values. The concept is outlined in the green box inset in Figure \ref{fig:2}a.  The network first maps the random vector $\mathbf{z}$ onto an $N$-by-$M$ matrix $A$.  These values can vary continuously and take any real number value.  A Softmax function is then applied to each row of $A$ to generate a probability matrix $P$: 
\begin{equation}
    P_{ij} = \frac{\exp{(\alpha \cdot A_{ij})}}{\sum_{j=1}^{M}\exp{(\alpha \cdot A_{ij})}}
\end{equation}
The $i^{th}$ row of matrix $P$ is a $1\times M$ vector and represents the probability distribution that the $i^{th}$ thin-film layer takes on a particular material choice within the material library.  We use the SoftMax function because it produces a properly normalized probability distribution and is commonly used in other related tasks, such as classification tasks \cite{bishop2006pattern}.   The expected refractive index of the $i^{th}$ layer given by this distribution, calculated as $n_i(\lambda)=\sum_{j=1}^M m_j(\lambda) \cdot P_{ij}$, is used to define the thin film stack in subsequent TMM calculations in Res-GLOnet.  All functions in this algorithm can be expanded into a differentiable computational graph, meaning that the loss function gradient with respect to the refractive index is able to backpropagate through the probability matrix $P$ and to the network weights $\phi$.  

$\alpha$ is a hyperparameter that tunes the sharpness of the Softmax function.  Initially, $\alpha$ is set to be one and the expected refractive index of the $i^{th}$ layer has contributions from many different materials in the material library.  Over the course of network training, $\alpha$ is gradually and manually increased to a point where the probability distribution of the $i^{th}$ thin-film layer is effectively a delta function that has converged to a single material.



\subsection{ResNet generator}

Our optimization problem involves searching within a highly complex, non-convex design space and is made particularly challenging by device requirements spanning a wide range of incident wavelengths and angles.  In the early and intermediate stages of network training, a deep neural network is required to properly generate a complex distribution of devices spanning large regions of the design space.  However, towards the latter stages of network training, the distribution of generated devices should ideally converge to a simple and narrow distribution centered around the global optimum, which is more ideally modeled using a shallow network.  GLOnet schemes that train using a fixed network architecture do not have the flexibility to capture these trends: deep architectures have general difficulty in training due to the well known vanishing gradient problem, while shallow architectures have the issue of underfitting the design space and are ineffective during the early and intermediate stages of network training \cite{he2016deep}.

To address these issues, we utilize deep residual networks for the generator architecture, which reformulates our algorithm as Res-GLOnets.  Residual networks \cite{he2016deep} were developed in the computer vision community to stably process images in very deep networks and overcome the vanishing gradient problem, with the insight that the use of skip connections can enable the depth of the network to be effectively and implicitly tuned over the course of training.  A schematic of our Res-GLOnet architecture is shown in the blue box inset in Figure \ref{fig:2}a and comprises a series of sixteen residual blocks.  Each block contains a fully connected layer, batch normalization layer, and a leaky ReLU nonlinear activation layer.  The input $x_{in}$ and output $x_{out}$ of each residual block have the same dimension, and the output of each block contains contributions from both the residual block $f(x_{in})$ and skip connection: $x_{out}=f(x_{in})+x_{in}$.

The evolution of the Res-GLOnet architecture over the course of network training is sketched in Figure \ref{fig:2}c.  When the network is training in the early and intermediate stages of the optimization process, each residual block outputs terms that are typically larger than the skip connection contributions.  As a result, the network architecture functions as a deep network, which is required during these stages of Res-GLOnets training. As network training progresses, some of the residual blocks start to output relatively small contributions and $x_{out} \approx x_{in}$, due to the emergence of vanishing gradients.  The network architecture now functions as a shallow architecture, having effectively skipped over some of the residual blocks.  Note that the increasing contribution of skip connections and reduction of network complexity is not explicitly and externally controlled but evolves over the course of network training, as the loss function guides the network output distribution to a relatively simple form.



\section{Optimization of an anti-reflection coating}

\begin{figure*}[ht!]
    \centering
    \includegraphics[width = \linewidth]{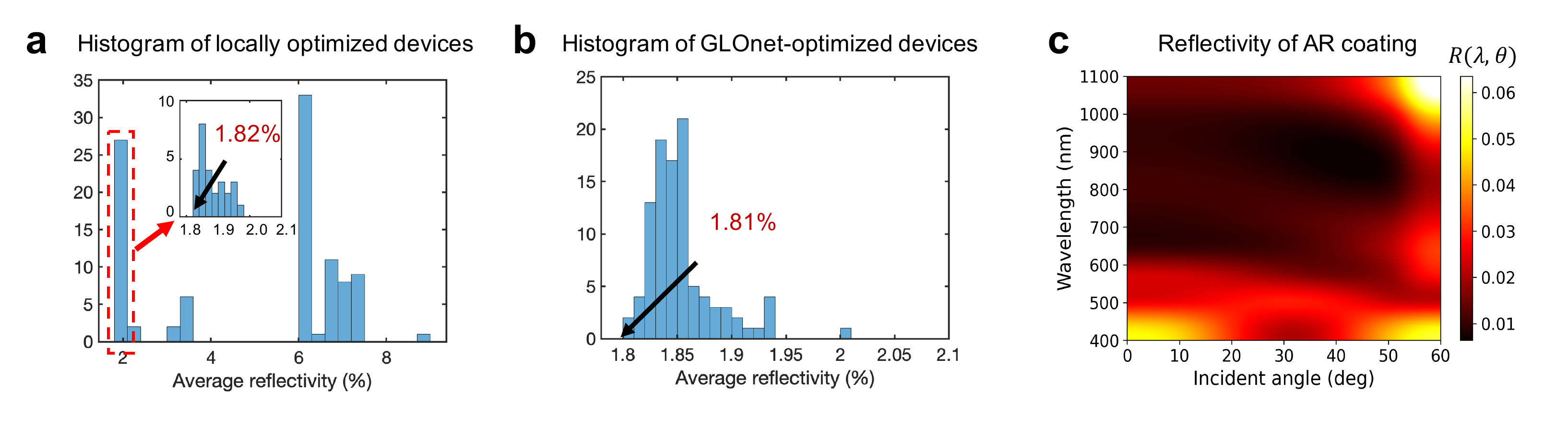}
    \caption{Optimization of a 3-layer thin film anti-reflection (AR) coating on silicon.  (a) Histogram of the average reflectivity from 100 AR coatings designed using local gradient-based optimization.  The best device has an average reflectivity of 1.82\%.  (b) Histogram of the average reflectivity from 100 AR coatings designed using a single Res-GLOnet. The best device has an average reflectivity of 1.81\%.  (c) Contour plot of reflectivity from the best Res-GLOnet-designed AR coating in (b) as a function in of incidence angle and wavelength, averaged for both TE- and TM-polarized waves.}
    \label{fig:3}
\end{figure*}

We first apply our Res-GLOnet algorithm to the design of a three layer anti-reflection (AR) coating for a silicon solar cell.  The thin-film AR stack is designed to minimize the average reflection at an air-silicon interface over the incident angle range [$0^{\circ}$, $60^{\circ}$] and wavelength range [400, 1100] nm for both TM and TE polarization.  As a benchmark, we compare our results with those from Ref. \cite{azunre2019guaranteed}, which provides a guaranteed global optimum solution using a parallel branch-and-bound method.  The algorithm requires extensive searching through the full design space and utilized over 19 days of CPU computation to solve for the global optimum.  To be consistent with Ref. \cite{azunre2019guaranteed}, the refractive indices of the layers in our design implementation do not take discrete categorical values from a materials library but are dispersionless and continuously varying in the interval [1.09, 2.60].  The thicknesses of each layer are also continuous variables within the interval [5, 200] nm.  

To accommodate the continuous variable nature of the refractive index values in this problem, we modify our categorical optimization scheme by setting the hyperparameter $\alpha = 1$ as a constant and specifying the material library to contain only two materials with constant refractive indices $\{m^L, m^U\}$.  $m^L = 1.09$ is the lower bound of the refractive index while $m^U = 2.60$ is the upper bound. The constraint on thickness can be satisfied by a transformation: $\mathbf{t} = t^L + \text{Sigmoid}(\tilde{\mathbf{t}}) \cdot (t^U - t^L)$. Here, the thickness directly outputted by the generator, $\tilde{\mathbf{t}}$, is normalized to [0, 1] and then linearly transformed to the interval [$t^L$, $t^U$], where $t^L = 5$ and $t^U = 200$ are the lower and upper thickness bound, respectively.

As a reference, we first optimize devices using local gradient-based optimization, by replacing the ResNet generator in our Res-GLOnet algorithm with an individual device layout.  The optimizations are performed with 100 different devices, initialized using random thickness and refractive index values within the limits of [1.09, 2.60] and [5, 200] nm, respectively.  Each optimization is performed over 200 iterations, so that a total of 20,000 sets of calculations is performed for the entire set of optimizations.  A histogram of the results (Figure \ref{fig:3}a) show that the optimized devices have average reflectivities that span a wide range of values, from approximately 2\% to 10\%, demonstrating the highly non-convex nature of the design space.  Average reflectivity is calculated  as the reflectivity averaged over the wavelengths, incident angles, and polarizations covered in the design specifications.  A fraction of devices are near the global optimum, and the best device has an efficiency of 1.82\%.  

A histogram of devices sampled from a single trained Res-GLOnet is summarized in Figure \ref{fig:3}a.  A total of 200 iterations is used together with a batch size of 20 devices, so that a total of 4,000 sets of calculations is performed.  The total time that Res-GLOnet requires for training is seven seconds with a single GPU.  All of the devices sampled from the Res-GLOnet are near the global optimum, showing the ability for the generative network to produce a narrow distribution of devices centered at the global optimum.  The best device has an efficiency of 1.81\% and its reflectivity for differing incident wavelengths and angles are plotted in Figure \ref{fig:3}c.  The design of this best device is summarized in Table \ref{tab:1} and is consistent with the result reported in Ref. \cite{azunre2019guaranteed}.  

\begin{table}[h!]
    \centering
    \begin{tabular}{@{}ccc@{}}
    \hline
       Layer \#  &  Refractive index & Thickness (nm) \\
    \hline
                &        Air        &   superstrate \\
        1       &       2.60        &      54.2     \\
        2       &       1.68        &      93.6     \\
        3       &       1.17        &      149.2    \\
                &       Si          &    substrate  \\
    \hline
    \end{tabular}
    \caption{Optimized structure for AR coating of Si}
    \label{tab:1}
\end{table}


\section{Optimization of incandescent light bulb filter}
\begin{figure*}[ht!]
    \centering
    \includegraphics[width = 0.9\linewidth]{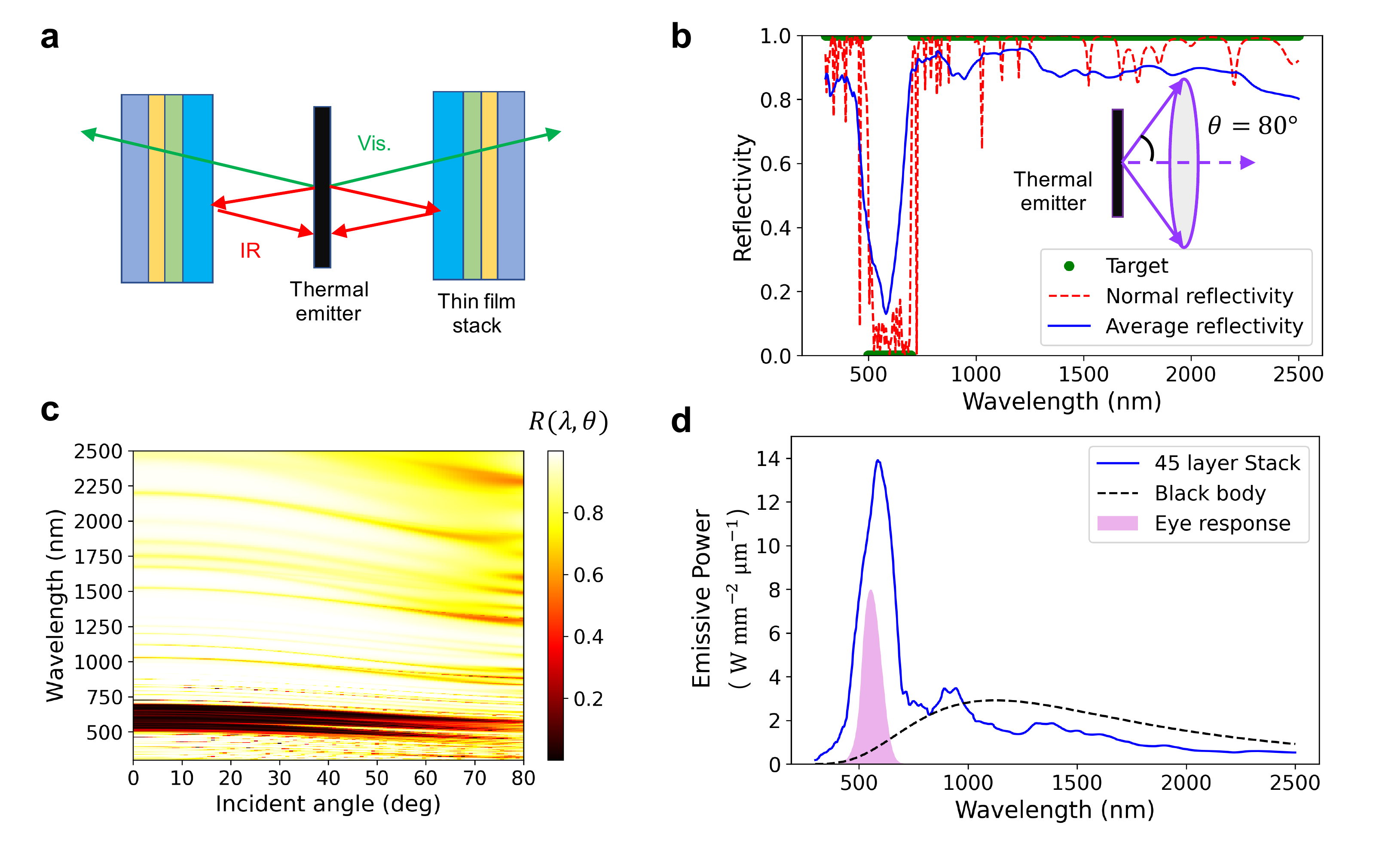}
    \caption{Thin film stacks for incandescent light bulb filtering. (a) Schematic of an incandescent light bulb filter that transmits visible light and reflects infrared and ultraviolet light.  (b) Reflection spectra of a 45-layer Res-GLOnet-optimized device, for normally incidence waves and waves averaged over a large incident solid angle, shown in the inset.  (c) Reflection spectra of the device featured in (b) as a function of incident angle, averaged for TE- and TM-polarized incident waves.  (d) Emissive power of a blackbody incandescent source and an equivalent source sandwiched by the filter featured in (b).  Also shown is the spectral response of the eye.}
    \label{fig:4}
\end{figure*}

To explore the applicability of Res-GLOnets to more complex problems, we apply our algorithm to optimize incandescent light bulb filters that transmit visible light and reflect infrared light (Figure \ref{fig:4}a). In this scheme, the emitter filament heats to a relatively higher temperature using recycled infrared light, thereby enhancing the emission efficiency in the visible range \cite{ilic2016tailoring}.  




A range of design methods have been previously applied to this problem.  In the initial demonstration of the concept, binary thin-film stacks were designed using a combination of local gradient-based optimization, used to tune the thickness of each layer, and needle optimization, which determined whether an existing layer should be removed or a new layer should be introduced \cite{ilic2016tailoring}.  A memetic algorithm was subsequently applied in which crossover, mutation, and downselecting operations were iteratively performed on a population of thin-film stacks to evolve the quality of devices \cite{shi2017optimization}.  Gradient-based local optimizations of device thicknesses were also periodically performed to refine the structures and accelerate algorithm convergence.  In a third study, reinforcement learning (RL) was used in which an auto-regressive recurrent neural network generated thin-film stacks layer-by-layer as a sequence \cite{wang2020automated}.  Unlike the GLOnet generator, the probability distribution of the thin-film stack was explictly outputted by the auto-regressive generator. The distribution evolved by optimizing a reward function, and the gradient of the reward function with respect to the neural network weights was calculated using proximal policy optimization.



In our demonstration, we benchmark Res-GLOnets with the memetic and RL studies, which consider a material library comprising seven dielectric material types: Al$_2$O$_3$, HfO$_2$, MgF$_2$, SiC, SiN, SiO$_2$ and TiO$_2$.  The superstrate and substrate are both set to be air. The complete wavelength range under consideration is [300, 2500] nm, and the target reflection is set to be 0\% for the wavelength range [500, 700] nm and 100\% for all other wavelengths. The incident angles span [0, 72] degrees and both TE and TM polarization are considered.

We train a Res-GLOnet comprising 16 residue blocks for 1000 iterations with a batch size of 1000.  The network is optimized using gradient decent with the momentum algorithm ADAM \cite{kingma2014adam}, and a learning rate of $1 \times 10^{-3}$ is used. The broadband reflection characteristics of a 45-layer device shows that the device operates with nearly ideal transmission in the [500, 700] nm interval and nearly ideal reflection at ultraviolet and near-infrared wavelengths, for both normal incidence and for incidence angles averaged over all solid angles within [0, 80] degrees (Figures \ref{fig:4}b and \ref{fig:4}c).  The emission intensity spectrum of the light bulb with and without the thin film filter are shown in Figure \ref{fig:4}d. The input power is fixed at 100 W and the surface area of the emitter is 20 mm$^2$.
\begin{figure}[ht!]
    \centering
    \includegraphics[width = 0.95\linewidth]{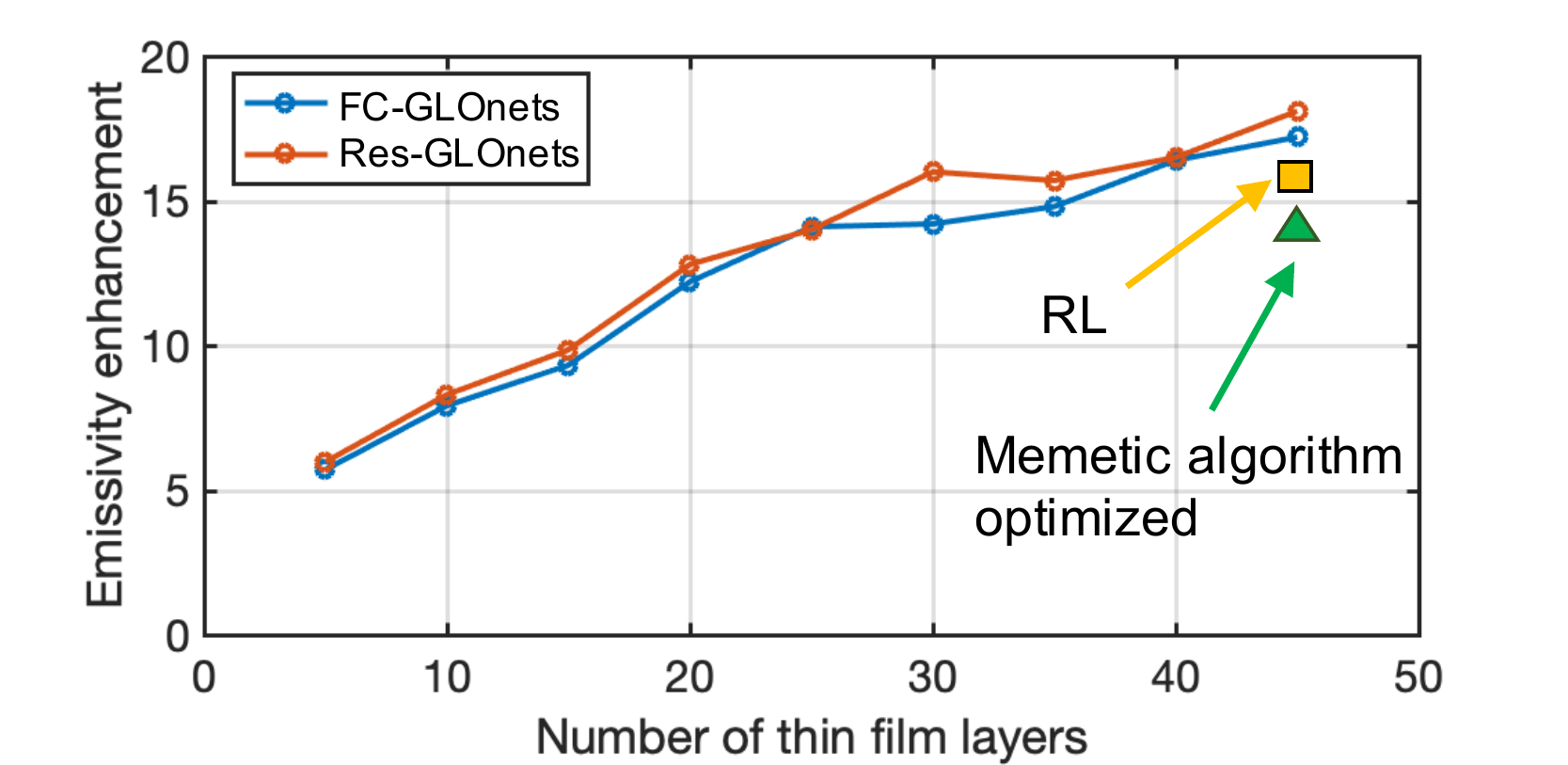}
    \caption{Plot of emissivity enhancement as a function of the number of thin film layers, for devices optimized using Res-GLOnets and FC-GLOnets.  Reference points are also plotted for devices designed using the RL \cite{wang2020automated} and memetic \cite{shi2017optimization} algorithm.}
    \label{fig:5}
\end{figure}

To evaluate the enhancement of visible light emission due to the filter, we compute the emissivity enhancement factor, $\chi$, as a function of the number of thin film layers:
\begin{equation}
    \chi = \frac{\int_0^{\infty} E_{\text{emitter+stack}}(P_0, \lambda) V(\lambda) d\lambda}
    {\int_0^{\infty} E_{\text{emitter}}(P_0, \lambda) V(\lambda) d\lambda}
\end{equation}
$E_{\text{emitter+stack}}(P_0, \lambda)$ and $E_{\text{emitter}}(P_0, \lambda)$ are the intensity emission spectrum given the input power $P_0$.  $V(\lambda)$ is the eye’s sensitivity spectrum and is shown as the shaded region in Figure \ref{fig:4}d.  The view factor is the proportion of emitted light from the light bulb filament that can reach the light bulb filter. We use the view factor of 0.95 as was the case for memetic study \cite{shi2017optimization}.  For a 45-layer device, the Res-GLOnet-optimized device achieved a $\chi$ of 17.2, and devices with as few as 30 layers still achieved a $\chi$ above 15 (Figure \ref{fig:5}). The ability to realize high performance devices with relatively few layers is practically important from a manufacturing and cost point of view.  The 45-layer memetic algorithm and RL-optimized device have $\chi$ values of 14.8 and 16.6, respectively. We also benchmark the Res-GLOnets with GLOnets based on a fixed architecture of four fully connected layers (FC-GLOnets). The benchmark, also plotted in Figure \ref{fig:5}, shows that Res-GLOnets performs better in searching for proper devices in this non-convex optimization problem, particularly for systems with larger numbers of thin films.  


\par
\section{Conclusion}

In summary, we show that Res-GLOnets are effective and efficient global optimizers for the multi-objective, categorical design of thin-film stacks.  Categorical optimization is performed through the use of a probability matrix, which is fully differentiable and compatible with our neural network training framework.  The incorporation of skip connections in our generative neural network helps it evolve from a deep to shallow architecture, which fits with our training objective and improves our search for the global optimum.  Benchmarks of our algorithm with known AR coating and incandescent light filter systems indicate that Res-GLOnets is effective at searching for global optima, is computationally efficient, and outperforms a number of alternative design algorithms.


We anticipate that concepts developed within Res-GLOnets, particularly those in categorical optimization, can directly apply to the design of other photonics systems, such as lens design where the material type is selected from a materials database.  We also expect that the implementation of application-specific electromagnetic solvers, in conjunction with automatic differentiation packages, will serve as a foundational concept for many high speed optimization algorithms beyond those for thin-film stacks.  Looking ahead, we see opportunities for Res-GLOnets to apply to other fields in the physical science, ranging from materials science and chemistry to mechanical engineering, where devices and systems are designed using combinations of discrete material types.


\appendix
\bibliography{refs}

\end{document}